\documentclass[aps,prr,floatfix,twocolumn,showpacs,preprintnumbers,amsmath,amssymb,superscriptaddress]{revtex4-2}
\usepackage[utf8]{inputenc}
\usepackage{amsmath}
\usepackage{amssymb}
\usepackage{graphicx}
\usepackage{bm}
\usepackage{color} % To highlight sentences when editing; remove before submission
\usepackage{relsize}
\usepackage{braket}
\usepackage{bbold}
\usepackage{epstopdf}
\usepackage{float}
\usepackage{placeins}
\usepackage[inkscapelatex=false]{svg}

\begin{document}
\title{Anisotropic Superconducting Diode Effect in Planar Josephson Junctions}
\author{A. Chilampankunnel Prasannan}
\affiliation{Department of Physics and Astronomy, Wayne State University, Detroit, MI 48201, USA}

\author{B. Pekerten}
\affiliation{Department of Physics, University at Buffalo, State University of New York, Buffalo, New York 14260, USA}

\author{N. Alashkar}
\affiliation{Department of Physics and Astronomy, Wayne State University, Detroit, MI 48201, USA}

\author{A. Matos-Abiague}
\affiliation{Department of Physics and Astronomy, Wayne State University, Detroit, MI 48201, USA}
\date{\today}

\begin{abstract}

We theoretically investigate the magnetic and crystalline anisotropies of the superconducting diode effect (SDE) in proximitized planar Josephson junctions (JJs) with coexisting Rashba and Dresselhaus spin–orbit couplings (SOCs) under an in-plane magnetic field. A symmetry analysis identifies geometric constraints on magnetic-field and crystallographic orientations for which the SDE is suppressed independently of field strength, providing experimentally testable signatures of the interplay between SOC and Zeeman interaction. We develop a phenomenological model showing that the diode efficiency depends on the relative alignment between spin–orbit and magnetic fields, and corroborate this behavior in the narrow-junction, low-field regime using an analytical approach that links the anisotropy of the diode response to SOC-induced Fermi surface distortions and anisotropic Cooper pair momentum. These findings are supported by tight-binding simulations of the Bogoliubov–de Gennes equation, which reproduce recent experimental trends. The simulations reveal that electrostatic gating can induce polarity reversals of the SDE in the low-field regime even with only Rashba SOC, consistent with recent experiments, and predict additional reversals for specific field orientations, junction geometries, and SOC ratios. Our results elucidate the origin of anisotropic nonreciprocal superconducting transport and provide guidance for experimentally probing the mechanisms underlying the SDE in semiconductor-based planar JJs.

\end{abstract}

\maketitle
\section{Introduction}

Nonreciprocal transport phenomena in superconductors have become a central topic in condensed matter physics due to their intimate connection with broken inversion and time-reversal symmetries and their potential applications in superconducting electronics. In particular, the superconducting diode effect (SDE), characterized by different critical currents depending on the direction of flow, has been investigated in a wide variety of material platforms and heterostructures \cite{Yokoyama2013:JPSJ,Yokoyama2014:PRB,Wakatsuki2018:PRL,Vodolazov2018:SST,Chen2018:PRB,Pal2019:EPL,Ando2020:N,Baumgartner2022:NN,Baumgartner2022:JPCM,Turini2022:NL,Pal2022:NP,Jeon2022:NM,Wu2022:N,Bauriedl2022:NC,Tanaka2022:PRB,Lin2022:NP,Hou2023:PRL,Costa2023:NN,Diez-Merida2023:NC,Zhao2023:S,Lu2023:PRL,Legg2023:PRB,Trahms2023:N,Lotfizadeh2024:CP,Reinhardt2024:NC,Ghosh2024:NM,Fu2024:PRApp,Cayao2024:PRB,Meyer2024:APL,Debnath2024:PRB,Banerjee2024:PRB,Zhang2024:NC,Scharf2024:PRB,zhong2025:arxiv}. These diverse realizations highlight the richness of microscopic mechanisms leading to supercurrent rectification \cite{Davydova2022:SA,Fracassi2024:APL,Shaffer2025:arxiv}.

In this work, we focus on semiconductor-based planar Josephson junctions (JJs), where the SDE has been robustly demonstrated experimentally \cite{Baumgartner2022:NN,Baumgartner2022:JPCM,Turini2022:NL,Costa2023:NN,Lotfizadeh2024:CP,Reinhardt2024:NC,Schiela2025b:arxiv}. A commonly invoked mechanism in these heterostructures is the formation of Cooper pairs with finite center-of-mass momentum under broken inversion and time-reversal symmetries, typically induced by Zeeman fields or exchange splitting \cite{Edelstein1989:JETP,Smith2021:PRB,Daido2022:PRL,Yuan2022:PNAS,He2022:NJP,Ilic2022:PRL,Davydova2022:SA,Baumgartner2022:NN,Baumgartner2022:JPCM,Fuchs2022:PRX,Costa2023:NN,Hou2023:PRL,Banerjee2023:PRL,Picoli2023:PRB,Sundaresh2023:NC,Lotfizadeh2024:CP,Reinhardt2024:NC}. However, the precise role of spin–orbit coupling (SOC) remains an open and actively debated question. While several theoretical works emphasize SOC-driven nonreciprocity \cite{Smith2021:PRB,Daido2022:PRL,Yuan2022:PNAS,He2022:NJP,Ilic2022:PRL,Baumgartner2022:NN,Baumgartner2022:JPCM,Costa2023:NN,Kochan2023:arXiv,Costa2023:PRB,Lotfizadeh2024:CP,Schiela2025b:arxiv,Pekerten2026:arxiv}, alternative approaches attribute the effect to mechanisms such as Doppler-shifted quasiparticle spectra, orbital contributions, edge reconstruction, diamagnetic effects induced by stray fields, among others \cite{Davydova2022:SA,Picoli2023:PRB,Hou2023:PRL,Sundaresh2023:NC,Banerjee2023:PRL,Fracassi2025:arxiv,Shaffer2025:arxiv}. Nevertheless, a wide range of experimental and theoretical studies across different JJ platforms \cite{Baumgartner2022:NN,Baumgartner2022:JPCM,Pal2022:NP,Costa2023:NN,Turini2022:NL,Mazur2024:PRA,Lotfizadeh2024:CP,Reinhardt2024:NC,Schiela2025b:arxiv,Yu2025:arxiv,Banerjee2023:PRL,He2022:NJP,Yuan2022:PNAS,Yokoyama2014:PRB,Yokoyama2013:JPSJ,Costa2023:PRB,Scharf2024:PRB,Meyer2024:APL,Pekerten2024b:APL,Pekerten2026:arxiv,Lu2023:PRL,Kochan2023:arXiv} consistently highlight the crucial role of magnetic fields and spin–orbit textures in determining both the magnitude and directionality of the diode response.

Semiconductor–superconductor heterostructures combine strong SOC with electrostatic tunability of carrier density and junction transparency, enabling systematic control of diode behavior through external gating \cite{Baumgartner2022:NN,Banerjee2023:PRL,Lotfizadeh2024:CP,Schiela2024:PRXQ,Schiela2025a:arxiv,Schiela2025b:arxiv,Lombardi2025:CM}. Recent experiments have further revealed a dependence of the diode effect on crystallographic orientation, indicating that crystalline anisotropy plays a key role in determining its magnitude and even its sign \cite{Baumgartner2022:JPCM,Schiela2025b:arxiv}. In junctions based on zinc-blende semiconductors, bulk inversion asymmetry gives rise to Dresselhaus SOC in addition to Rashba SOC induced by structural inversion asymmetry. This interplay introduces a strong dependence on crystal axes, which can significantly reshape transport properties and, consequently, the behavior of the SDE \cite{Baumgartner2022:JPCM,Schiela2025b:arxiv}.

In this work, we investigate the magnetic and crystalline anisotropies of the SDE in semiconductor-based planar JJs (such as Al/InAs and Al/InSb) hosting coexisting Rashba and Dresselhaus SOC under an in-plane magnetic field. Using a symmetry-based analysis, we identify geometric conditions involving the magnetic-field and junction orientations under which the diode effect vanishes independently of the field strength, thereby providing robust and experimentally accessible signatures of the role of SOC in the emergence of the SDE. We further develop a phenomenological framework demonstrating that the diode efficiency is governed by the relative alignment between the effective spin–orbit field and the applied magnetic field. To elucidate this connection, we derive an analytical description in the narrow-junction, low-field regime, where the diode response can be directly traced to SOC-induced distortions of the Fermi contours and the resulting anisotropy in the Cooper-pair momentum. These analytical results are corroborated by tight-binding simulations of the Bogoliubov–de Gennes equation, which reproduce key experimental trends and capture the full angular dependence of the diode response. Moreover, the simulations reveal that electrostatic gating can induce reversals of the diode polarity in the low-field regime \cite{Schiela2025b:arxiv}, even in the presence of purely Rashba SOC.

Our results establish a framework for understanding how the interplay between SOC and Zeeman coupling gives rise to anisotropic nonreciprocal superconducting transport in planar JJs, and provide clear guidance for experimental probes of the role of SOC in the SDE in semiconductor-based Josephson junctions.

\section{Theoretical Model}

\begin{figure}[t!]
    \centering
    \includegraphics*[width=1\columnwidth]{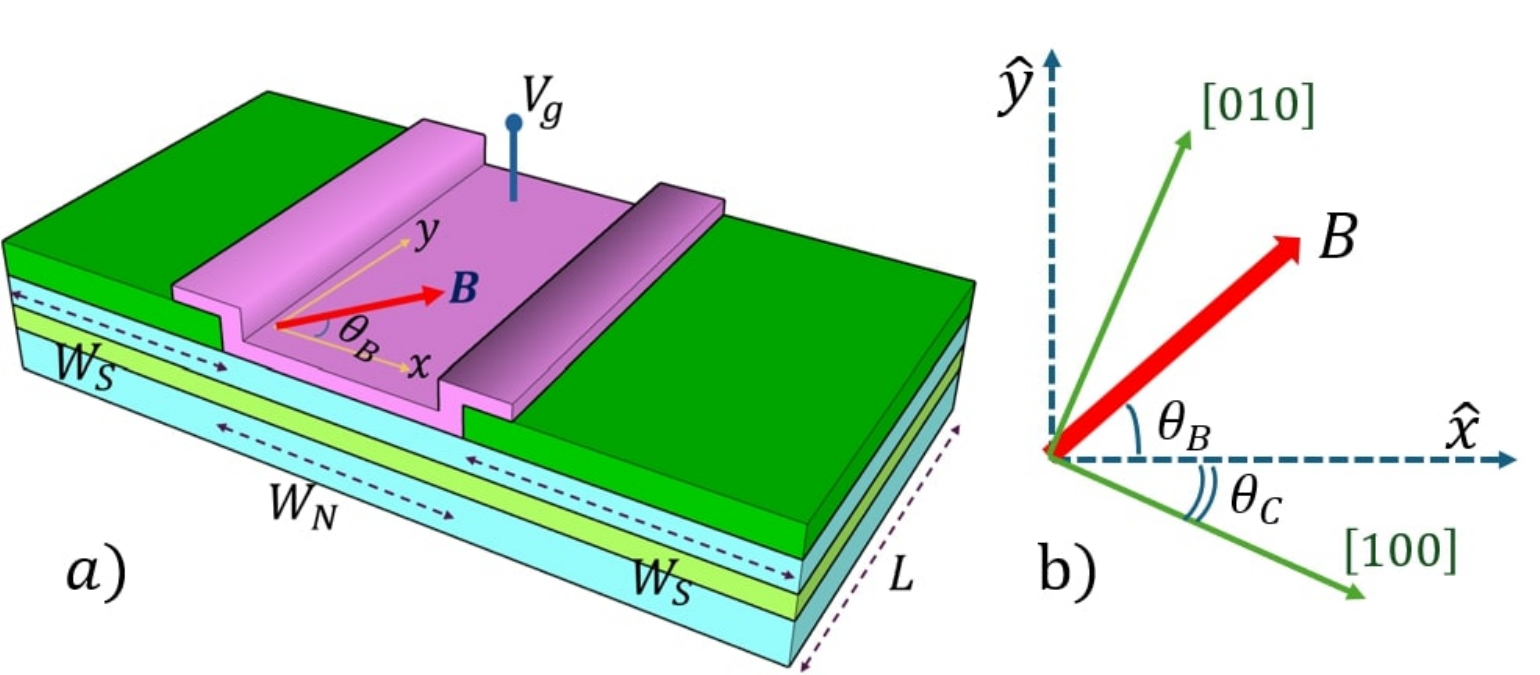}
    \caption{(a) Schematic of a gated planar JJ formed from a semiconductor two-dimensional electron gas (2DEG, light green). Superconducting leads (dark green) induce superconductivity in the 2DEG via the proximity effect. The junction transparency is controlled by a top-gate voltage $V_g$ applied to the normal region. The normal (N) region has width $W_N$, while each superconducting (S) lead has width $W_S$.
(b) Definition of the angles $\theta_B$ and $\theta_C$, corresponding to the orientations of the magnetic field $\mathbf{B}$ and the crystal axis $[100]$, respectively, measured from the current direction $\hat{x}$.}\label{fig:system}
\end{figure}

We consider a planar JJ built on a proximitized superconducting two-dimensional electron gas (2DEG), as depicted in Fig.~\ref{fig:system}(a). The junction is subject to an in-plane magnetic field $\mathbf{B}$ oriented at an angle $\theta_B$ with respect to the current direction $\hat{x}$. The crystallographic orientation of the junction is such that the current direction (the $\hat{x}$ axis) forms an angle $\theta_C$ with the $[100]$ crystallographic axis [Fig.~\ref{fig:system}(b)].

The Bogoliubov–de Gennes (BdG) Hamiltonian describing the considered system reads
\begin{equation}\label{BdG}
    H =\sigma_z \otimes (H_0+H_{\text{SO}}) + H_Z + \Delta(x,\phi) \tau_+ + \Delta^*(x,\phi)\tau_- ,
\end{equation}
where the single-particle contribution $H_0 + H_{\rm SO}$ comprises the term
\begin{equation}\label{freeparticle}
    H_0 = \Big[ \frac{\mathbf{p}^2}{2m^*}-\mu_S+V(x) \Big]\sigma_0\;,
\end{equation}
and the SOC Hamiltonian $H_{\text{SO}} = H_R + H_D$. In zinc-blende–based quantum-well structures, the SOC consists of the superposition of Rashba SOC \cite{Rashba1984:JETP},
\begin{equation}\label{Rashba}
    H_R = \frac{\alpha}{\hbar}(p_y\sigma_x-p_x\sigma_y)\;,
\end{equation}
which originates from the structural inversion asymmetry of the heterostructure, and the linearized Dresselhaus SOC \cite{Zutic2004:RMP,Fabian2007:APS},
\begin{equation}\label{Dresselhaus}
    H_D = \frac{\beta}{\hbar} \big[(p_x\sigma_x-p_y\sigma_y)\cos2\theta_C-(p_x\sigma_y+p_y\sigma_x)\sin2\theta_C\big],
\end{equation}
which results from the lack of bulk inversion symmetry \cite{Dresselhaus1955:PR}.
Note that we consider only the linear SOC contributions. According to the results reported in this study, these terms are sufficient to qualitatively capture the general trends observed in previous experiments on the SDE in planar JJs \cite{Baumgartner2022:JPCM,Costa2023:NN,Lotfizadeh2024:CP}. Nevertheless, cubic SOC contributions may become relevant in certain situations \cite{Pekerten2024b:APL,Schiela2025b:arxiv}.

In Eqs.~(\ref{freeparticle})–(\ref{Dresselhaus}), $p_x$ and $p_y$ denote the in-plane momentum components, $m^*$ is the effective mass, and $\theta_C$ is the angle between the $[100]$ crystallographic direction and the $x$ axis. The potential
\begin{equation}\label{V(x)}
V(x) = V_0 \Theta(W_N/2-|x|) \; , V_0  = \mu_S - \mu_N\; .
\end{equation}
with $\Theta(x)$ as the Heaviside function, describes the difference between the chemical potentials in the superconducting (S) and normal (N) regions, denoted by $\mu_S$ and $\mu_N$, respectively. The strengths of the Rashba and Dresselhaus SOC are denoted by $\alpha$ and $\beta$, respectively. The matrices $\sigma_{x,y,z}$ are the Pauli matrices, and $\sigma_0$ is the $2\times2$ identity matrix. We adopt the following SOC parametrization,
\begin{equation}\label{SOCparameterization}
    \alpha = \lambda \cos\theta_{so},\;\beta=\lambda \sin\theta_{so}\;,
\end{equation}
where $\lambda$ represents the overall strength of the combined Rashba and Dresselhaus SOCs, and $\theta_{so}$ is the SOC angle characterizing the relative contribution of the two couplings. Explicitly,
\begin{equation}
    \lambda = \sqrt{\alpha^2+\beta^2},\;\theta_{so}=\arctan(\beta/\alpha)\;.
\end{equation}

The third term in Eq.(~\ref{BdG}) accounts for the Zeeman interaction,
\begin{equation}\label{Zeeman}
    H_Z = - \mathbf{E_Z} \cdot (\sigma_0 \otimes \mathbf\sigma) \;,
\end{equation}
due to the in-plane magnetic field,

  \begin{align}
    \mathbf{E_Z} &= \frac{g^*\mu_B}{2}|\mathbf{B}|\begin{pmatrix}
           \cos\theta_B \\
           \sin\theta_B \\
           0
         \end{pmatrix}\;.
  \end{align}
  
Here, $\mu_B$ as the Bohr magneton and $g^*$ is the effective $g$-factor. The last two terms of Eq. (\ref{BdG}) represent the superconducting pairing potential with $\Delta(x,\phi) = \Delta_0 e^{i\; \text{sgn}(x)\phi/2}\Theta(|x|-W_N/2)$ and $\tau_\pm=(\sigma_x \pm i\sigma_y)\otimes \sigma_0/2$. Here $\Delta_0$ is the proximity-induced superconducting gap and $\phi$ is the superconducting phase difference across the junction.

Solving the BdG equation yields the quasiparticle energy spectrum $\{E_{n,k_y}\}$. In the zero-temperature limit, the phase-dependent component of the free energy is given by,
\begin{equation}
    F(\mathbf{B},\phi)=\sum_{n,k_y} E_{n,k_y}^< \;,
\end{equation}
where $E_{n,k_y}^<$ denotes the negative eigenenergies of the BdG spectrum, with $n$ denoting the channel and $k_y$ as the wave vector along the junction direction. The free energy allows us to determine the current phase relation (CPR), 
\begin{equation}\label{CPR}
    I(\mathbf{B},\phi)=\frac{2 e}{\hbar} \frac{dF(\mathbf{B},\phi)}{d\phi}\;,
\end{equation}
where $e$ is the electron charge. The forward and reverse supercurrents flow along and against the gradient of the superconducting phase across the junction.

In phase-biased circuits, the junction is embedded in a loop threaded by a magnetic flux, which controls the superconducting phase difference. In such systems, the \emph{phase-biased} SDE is characterized by the supercurrent difference,
\begin{equation}
    \Delta I(\mathbf{B},\phi)=I(\mathbf{B},\phi)-|I(\mathbf{B},-\phi)|.
\end{equation}
On the other hand, in phase-unbiased junctions, where no external magnetic flux is applied, the superconducting phase difference self-adjusts to maximize the supercurrent amplitude. Within the phase convention adopted here, the forward ($I_c^{+}$) and reverse ($I_c^{-}$) critical currents are given by
\begin{eqnarray}\label{critical currents}
    I_c^+(\mathbf{B}) & = & \max_\phi I(\mathbf{B},\phi)\;,
    I_c^-(\mathbf{B})  =  \min_\phi I(\mathbf{B},\phi)\;.
\end{eqnarray}
The corresponding diode efficiency, which characterizes the strength of the \emph{phase-unbiased} SDE, is then introduced as
\begin{equation}\label{qualityfactor}
    \eta = \frac{|I_c^+(\mathbf{B})|-|I_c^-(\mathbf{B})|}{I_0}\;,
\end{equation}
where $I_0$ denotes the critical current at zero magnetic field.

While alternative definitions of the diode efficiency exist, the definition in Eq.~(\ref{qualityfactor}) is particularly convenient for analyzing anisotropic effects. The interplay between SOC and the Zeeman field leads to a magnetoanisotropy of the critical currents with respect to the magnetic-field orientation. Moreover, Dresselhaus SOC, which unlike Rashba SOC is not invariant under in-plane rotations, induces a crystalline anisotropy of the SDE, making the forward and reverse critical currents dependent on their direction relative to the junction’s crystallographic axes. Consequently, in the presence of Dresselhaus SOC, the diode quality factor exhibits both magneto- and crystalline anisotropies. Therefore, the experimental observation of these anisotropies in the SDE can serve as evidence of the interplay between SOC and the Zeeman interaction as the underlying mechanism behind the SDE, whereas the absence of SOC-induced anisotropic behavior would instead favor alternative mechanisms.

%%%%%%%%%%%%%%%%%%%%%%%%%%%%%%%%%%%%%%%%%%%%%%%%%%%%%%%%%%%%%%%%%
        % Symmetry requirements for vanishing SDE               % 
%%%%%%%%%%%%%%%%%%%%%%%%%%%%%%%%%%%%%%%%%%%%%%%%%%%%%%%%%%%%%%%%%
\begin{table*}[ht]
    \centering
    \setlength{\tabcolsep}{8pt}
    \renewcommand{\arraystretch}{1.50}
    \caption{Symmetry conditions for SDE suppression.} 
    \begin{tabular}{ccccccc}
    \hline \hline
    
       & $\alpha$               & $\beta$   &$\{k,l,m,n\}$  &  $\mathbf{u}^T$           & condition \\ \hline \hline  
        
       (1)&  $\neq 0$                & 0    &$\{1,0,1,0\}$    &   $(1,0,0)$              &$\sin\theta_B=0$\\

        \hline
         
       (2) &  0  &      $\neq 0$      & $\{1,0,1,0\}$  &    $(\sin 2\theta_C,\cos 2\theta_C,0)$   &  $\cos(2\theta_C+\theta_B)=0$\\

         \hline

       (3) &  $\neq 0$  &  $\neq 0\; \& \;\neq\alpha$  & $\{1,0,1,0\}$  &   $(1,0,0)$   &  $\cos2\theta_C=\sin\theta_B=0$\\

         \hline
                
       (4) & $\neq 0$        & $\pm \alpha$      & \{0,0,1,1\} &  $(\pm\alpha\cos 2\theta_C,-\alpha(1\pm\sin 2\theta_C),0)$            & $\cos(2\theta_C+\theta_B)=\pm\sin\theta_B$ \\

        \hline
        
       (5) & $\neq 0$        & $\pm \alpha$      &  &        & $\sin(2\theta_C+\theta_B)=\mp\cos\theta_B$ \\
        
        \hline \hline
    \end{tabular}\label{tab-1}
\end{table*}

\section{Symmetry analysis}

The planar JJs considered here, with SOC and Zeeman interaction, as described by the BdG Hamiltonian in Eq.~(\ref{BdG}) lack both time-reversal and space-inversion symmetries, which are general requirements for the emergence of superconducting nonreciprocal effects such as the SDE. Below, we perform a symmetry analysis revealing how some symmetry constraints in the studied JJs may lead to the suppression of the SDE.

The phase-biased SDE occurs when the amplitudes of the forward [$I(B,\phi)$] and reverse [$-I(B,-\phi)$] supercurrents differ. According to Eq.~(\ref{CPR}), this requires that $F(B,\phi)\neq F(B,-\phi)$. Therefore, the condition $E_n(B,\phi)=E_n(B,-\phi)$ constitutes a sufficient criterion for the suppression of the phase-biased SDE. This condition is equivalent to the existence of a unitary (or antiunitary) transformation $\mathcal{U}$ such that
\begin{equation}\label{u-sym}
\mathcal{U}^\dagger H(B,\phi)\;\mathcal{U}=H(B,-\phi).
\end{equation}

Since the SDE is associated with broken space-inversion and time-reversal symmetries, we consider transformations of the form
\begin{equation}\label{u-def}
\mathcal{U}_{klm}^{n}(\mathbf{u}) = R_x^{k} R_y^{l} \left[\Gamma(\mathbf{u})\right]^{m} T^{n},
\end{equation}
where $R_x$ and $R_y$ denote the spatial mirror operators with respect to the $yz$ and $xz$ planes, respectively,
\begin{equation}\label{gamma-def}
\Gamma(\mathbf{u}) = i\sigma_0 \otimes \left(\frac{\mathbf{u}}{|\mathbf{u}|} \cdot \boldsymbol{\sigma}\right),
\end{equation}
represents a spin reflection through the plane perpendicular to the vector $\mathbf{u}$, and $T=-i(\sigma_0\otimes\sigma_y)\mathcal{K}$ (with $\mathcal{K}$ the complex-conjugation operator) is the time-reversal operator. The exponents ${k,l,m,n}$ in Eq.~(\ref{u-def}) take values 0 or 1.

After some algebra, we identify a set of conditions under which Eq.~(\ref{u-sym}) is satisfied, leading to the suppression of the phase-biased SDE [see conditions (1)-(4) in Table \ref{tab-1}]. Additional constraints arise in the special case $\beta=\pm\alpha$. Under suitable geometric conditions, the SOC can be completely gauged away via a unitary transformation of the form
\begin{equation}\label{u-transf}
U = e^{i\mathbf{q}\cdot\mathbf{r}[\sigma_0 \otimes (\mathbf{v}\cdot\boldsymbol{\sigma})]},
\end{equation}
which consequently eliminates the SDE.

As a representative example, for $\alpha=\beta$, $\theta_C=\pi/4$, and $\theta_B=\pi/2$, the SOC can be removed by choosing $\mathbf{q} = (2m^\ast\alpha/\hbar^2, 0, 0)^T$ and $\mathbf{v} = (0,1,0)^T$ in Eq.~(\ref{u-transf}). More generally, one finds that the SOC can be gauged out—and hence the SDE suppressed—whenever $\beta=\pm\alpha\neq 0$ and the geometric constraint $\sin(2\theta_C+\theta_B) = \mp \sin\theta_B$ is satisfied [see condition (5) Table \ref{tab-1}].

The symmetry-imposed conditions for SDE suppression summarized in Table~\ref{tab-1} are determined solely by the symmetry of the SOC and the orientations of the magnetic field and the junction. Therefore, the conditions in Table~\ref{tab-1} can be used to experimentally probe whether the interplay between spin–orbit coupling and Zeeman interaction is indeed the macroscopic mechanism responsible for the SDE in semiconductor-based planar JJs.

\section{Phenomenological model}
To better understand the magnetic and crystalline anisotropies of the supercurrents, we formulate a general phenomenological model capable of providing insight into the interrelation between the relevant orientations ($\theta_B$ and $\theta_C$) and the superconducting phase difference, $\phi$.

The free energy of the system is a periodic function of the superconducting phase difference and can be written as a trigonometric series,
\begin{eqnarray}\label{f-trig}
    F(\mathbf{B},\phi)&=& F_0(\mathbf{B})+
    \sum_{m=1}^{\infty}a_m(\mathbf{B})\sin(m\phi)\nonumber\\
    &+&\sum_{m=1}^{\infty}b_m(\mathbf{B})\cos(m\phi),
\end{eqnarray}
where the expansion coefficients, according to the symmetry relation $F(\mathbf{B},\phi)=F(-\mathbf{B},-\phi)$, obey
\begin{gather}
     F_0(\mathbf{B})=F_0(-\mathbf{B}),\nonumber\\
     a_m(\mathbf{B})=-a_m(-\mathbf{B}),\; b_m(\mathbf{B})=b_m(-\mathbf{B}).
     \label{parity-constraints}
\end{gather}
It follows from Eq.~(\ref{f-trig}) that, since the SDE requires $F(\mathbf{B},\phi)\neq F(\mathbf{B},-\phi)$, the expansion coefficients $a_n(\mathbf{B})$ are the ones responsible for the phase-biased SDE.

In what follows, we restrict our analysis to cases where the SOC cannot be gauged out, allowing the coefficients $a_m$ and $b_m$ to be expanded in powers of the magnetic field $\mathbf{B}$ and the SOC field components,
\begin{equation}
    \mathbf{w}_{x} = \begin{pmatrix}
        \beta \cos2\theta_C \\
        -\beta \sin2\theta_C - \alpha \\
        0
    \end{pmatrix},
\end{equation}
and
\begin{equation}
    \mathbf{w}_{y} = \begin{pmatrix}
        \alpha - \beta \sin2\theta_C \\
        -\beta \cos2\theta_C \\
        0
    \end{pmatrix}.
\end{equation}
Note that these vectors define the SOC Hamiltonian $H_{\text{SO}} = H_R + H_D = \mathbf{W}_{so} \cdot \boldsymbol{\sigma}$, where the effective spin–orbit field is given by $\mathbf{W}_{so}=(p_x \mathbf{w}_x+\hbar k_y\mathbf{w}_y)/\hbar$. Taking into account the parity constraints in 
Eq.~(\ref{parity-constraints}), $a_m$ and $b_m$ can generally be expanded as,
\begin{widetext}
\begin{eqnarray}
    a_m&=&\sum_{n,k_y}\left[c_{mnk_y}^{(1,1)}\left(\langle\mathbf{W}_{so}\rangle_n\cdot\mathbf{B}\right)+c_{mnk_y}^{(3,3)}\left(\langle\mathbf{W}_{so}\rangle_n\cdot\mathbf{B}\right)^3+...\right],\nonumber\\
    b_m&=&\sum_{n,k_y}\left[d_{mnk_y}^{(0,0)}+d_{mnk_y}^{(2,0)}|\langle\mathbf{W}_{so}\rangle_n|^2+d_{mnk_y}^{(0,2)}|\mathbf{B}|^2+ d_{mnk_y}^{(2,2)}\left(\langle\mathbf{W}_{so}\rangle_n\cdot\mathbf{B}\right)^2+...\right],
\end{eqnarray}
\end{widetext}
where $\langle \cdots \rangle_n$ denotes the expectation value in the $n$th channel. Restricting to leading order in the magnetic field and spin–orbit coupling, one obtains the approximations
\begin{equation}\label{f-g-def}
a_m \approx f_m (\mathbf{w}_x \cdot \mathbf{B})\;\;,\;\;b_m \approx g_m,
\end{equation}
where $f_m$ and $g_m$ are coefficients absorbing the sums over $n$ and $k_y$. Note that the term involving $k_y\mathbf{w}_y$ is odd in $k_y$ and therefore vanishes upon summation over $k_y$. In addition, this summation can induce a spin–orbit dependence in the coefficients $a_m$ and $b_m$, arising from the mismatch between the Fermi wave vectors of the two spin channels (see Sec.~\ref{sec:analytical} for explicit calculations). Using Eqs.~(\ref{CPR}), (\ref{f-trig}), and (\ref{f-g-def}), the generic form of the supercurrent up to leading order in the magnetic and spin-orbit fields can be written as
\begin{equation}\label{i-fenom-0}
    I\approx\sum_{m=1}^{\infty}\left[g_m\sin(m\phi)-f_m(\mathbf{w}_x\cdot\mathbf{B})\cos(m\phi)\right],
\end{equation}
where, for brevity, the factor $-2 e m/\hbar$ has been absorbed into the coefficients $f_m$ and $g_m$. It then follows from Eq.~(\ref{i-fenom-0}) that, except in the special cases where the SOC can be gauged out, the SDE is suppressed when
\begin{equation}\label{wb-noSDE}
    \mathbf{w}_x\cdot\mathbf{B}=B\left[\beta\cos(2\theta_C+\theta_B)-\alpha\sin\theta_B\right]=0,
\end{equation}
a condition that is consistent with the corresponding symmetry constraints listed in Table \ref{tab-1}.

Using trigonometric identities, Eq.~(\ref{i-fenom-0}) can be recast as,
\begin{equation}\label{i-fenom-1}
    I\approx\sum_{m=1}^{\infty}I_m\sin(m\phi-\phi_m),
\end{equation}
where,
\begin{equation}
    I_m=\sqrt{g_m^2+f_m^2(\mathbf{w}_x\cdot\mathbf{B})^2},
\end{equation}
and
\begin{equation}
    \phi_m=\arctan\left[\frac{f_m(\mathbf{w}_x\cdot\mathbf{B})}{g_m}\right]
\end{equation}
denote, respectively, the amplitude and anomalous phase associated with the $m$th harmonic.

Note that the presence of anomalous phases ensures that $I(\mathbf{B},\phi)\neq -I(\mathbf{B},-\phi)$, thereby yielding a finite phase-biased SDE even when only a single harmonic is present. Conversely, if the supercurrent is dominated by a single harmonic (say, the $m$th harmonic), the forward and reverse critical currents share the same amplitude, $I_{m}$, and the phase-unbiased SDE vanishes. Therefore, a finite phase-unbiased SDE requires contributions from at least two harmonics in the supercurrent. A minimal phenomenological model for the SDE, retaining only the two lowest harmonics, is given by \cite{Baumgartner2022:NN,Costa2023:NN,Pal2022:NP},
\begin{equation}\label{i-2-harms}
    I\approx I_1\sin(\phi-\phi_1)+I_2\sin(2\phi-\phi_2).
\end{equation}
In the limit $I_1\gg I_2$, the phases that maximize and minimize the supercurrent are approximately $\phi_{max}\approx \pi/2+\phi_1$ and $\phi_{min}\approx -\pi/2+\phi_1$, respectively. In the weak magnetic-field regime, the diode efficiency then reduces to
\begin{equation}\label{eta-minimal}
    \eta\approx -2\sin(2\phi_2-\phi_1)\approx\frac{2f_2g_1-4f_1g_2}{g_1 g_2}(\mathbf{w}_x\cdot\mathbf{B}).
\end{equation}
This relation highlights the crystalline and magnetic anisotropies of the SDE, as reflected in its dependence on both the spin–orbit field (hence on the crystallographic orientation of the junction) and the direction of the applied magnetic field. As follows from Eq.~(\ref{eta-minimal}), the SDE is expected to vanish when the condition in Eq.~(\ref{wb-noSDE}) is satisfied.

It is worth emphasizing that Eq.~(\ref{eta-minimal}) represents a lowest-order approximation for the SDE and may fail to capture its behavior in certain systems, in particular perfectly transparent junctions, where the contribution to the diode efficiency linear in the magnetic field amplitude $B$ has been shown to vanish \cite{Pekerten2026:arxiv}. Within the minimal model [Eq.~(\ref{i-2-harms})], quadratic-in-$B$ contributions to $\eta$ can arise by relaxing the condition 
$I_1\gg I_2$ and/or by employing more accurate expressions for the critical phases $\phi_{max}$ and $\phi_{min}$.

\section{Analytical approximation}
\label{sec:analytical}

An analytical description of the magnetic and crystalline anisotropies of the SDE in planar JJs is, in general, quite involved. However, the problem simplifies considerably in the narrow-junction limit, defined by $W_N\ll\xi\ll W_S$, where $\xi$ is the superconducting coherence length. In this regime, the junction potential can be approximated by a Dirac delta function, $V(x) \approx V_0 W_N \delta(x)$, while the superconducting leads may be treated as infinitely wide, $W_S\rightarrow\infty$. The strength of the junction potential, $V_0$, determines the junction transparency,
\begin{equation}\label{Transparancy}
    \tau = \frac{1}{1+(\frac{V_0}{2 E_T})^2}\;,
\end{equation}
where $E_T=\pi \hbar v_{F}/(2 W_N)$ denotes the Thouless energy and $v_F $ is the Fermi velocity.

We first focus on the case where only Rashba SOC is present. At $k_y=0$, the corresponding spin–orbit field is oriented along the $y$-direction. In the low magnetic-field limit, the spin orientation is dominated by this spin–orbit field, such that only the $y$-component of the external magnetic field plays a significant role. Within these approximations, the scattering problem associated with the Bogoliubov–de Gennes (BdG) Hamiltonian can be solved analytically. The resulting Andreev bound-state energy spectrum is given by \cite{Scharf2019:PRB},
\begin{widetext}
\begin{eqnarray}\label{Energy}
    E_{\nu,\sigma}(k_y=0,\theta_B,\phi)\approx\Delta \frac{\text{sgn}(A^\sigma_\nu(V_0,E_Z,\theta_B,\phi))|C(V_0,E_Z,\theta_B,\phi)|}{\sqrt{A^\sigma_\nu(V_0,E_Z,\theta_B,\phi)^2+C(V_0,E_Z,\theta_B,\phi)^2}}\;,
\end{eqnarray}
where,
\begin{eqnarray}
    A^\sigma_\nu(V_0,E_Z,\theta_B,\phi) &=&2\pi\sigma{E_Z}\sin\theta_B+2\nu\sqrt{E_T^2\sin^2\phi+\pi^2\Big({E_Z}^2\sin^2\theta_B\cos^2\frac{\phi}{2}+V_0^2\sin^2\frac{\phi}{2}\Big)}\;,\quad \nu,\sigma= \pm 1\\
    C(V_0,E_Z,\theta_B,\phi)
    &=& \sqrt{4 E_T^2 \cos^2\frac{\phi}{2}+\pi^2(V_0^2-{E_Z}^2\sin^2\theta_B)}.
\end{eqnarray}
\end{widetext}
Here ${E_Z} = g^*\mu_B B/2$ denotes the Zeeman energy. The index $\sigma=\pm 1$ labels the two spin branches, while $\nu=\pm 1$ distinguishes the particle-like and hole-like solutions.

As shown in Ref.~\cite{Pekerten2026:arxiv}, the zero-temperature free energy in the low magnetic-field and weak SOC regime($k_{so}\ll k_F$) can be approximated as follows,
\begin{equation}\label{Free_energy}
    F\approx\frac{L}{\pi}\sum_\nu E_{\nu +}^<(k_y=0)k_{F+}+E_{\nu -}^<(k_y=0)k_{F-}\;,
\end{equation}
where $L$ is the junction length,
\begin{equation}\label{kf-pm}
    k_{F\pm}=k_F\pm k_{so},
\end{equation}
represent the intersections of the spin-split Fermi contours with the $k_y$-axis, $k_{F}=\sqrt{2m^\ast\mu/\hbar^2}$ is the Fermi wave vector, and $k_{so}= m^*\alpha/\hbar^2$.

The approximate free energy in Eq.~(\ref{Free_energy}) can then be used to evaluate the forward and reverse critical currents, as well as the corresponding diode efficiency. In particular, in the high-transparency limit ($1-\tau \ll 1$), one can expand the Andreev energy spectrum in powers of $V_0$. Keeping only terms in the leading order of $v_0$, $B$, and SOC, one obtains a simple relation for the diode efficiency \cite{Pekerten2026:arxiv},
\begin{equation}
    \eta\approx (1-\tau)^{1/4}\frac{k_{so}}{k_F}\phi_q
\end{equation}
where
\begin{equation}
    \phi_q = q W_N
\end{equation}
is the phase shift induced by the magnetic field,
\begin{equation}\label{q-cooper-0}
    q\approx\frac{\kappa_B^2}{k_F}\sin\theta_B\;\;,\;\;\kappa_B=\frac{\sqrt{2m^\ast E_Z}}{\hbar^2}.
\end{equation}
is the Cooper pair momentum along the current direction, and 
\begin{equation}
   \kappa_B=\frac{\sqrt{2m^\ast E_Z}}{\hbar^2}.
\end{equation}

Rashba spin–orbit coupling (SOC) is invariant under in-plane rotations and therefore cannot, by itself, give rise to crystalline anisotropy; it can only induce magnetic anisotropy. This is reflected in Eqs.~(\ref{Free_energy})–(\ref{q-cooper-0}), which are independent of the crystallographic orientation of the junction ($\theta_C$), while exhibiting a dependence on the orientation of the applied magnetic field ($\theta_B$). In contrast, the inclusion of Dresselhaus SOC breaks the in-plane rotational invariance, thereby giving rise to both crystalline and magnetic anisotropies.

We now extend the analysis to junctions in which both Rashba and Dresselhaus spin–orbit couplings (SOC) are present. In this case, it is convenient to apply the unitary transformation
 \begin{eqnarray}
 U_{\theta} &=& e^{-i\frac{\theta}{2}\sigma_z\otimes\sigma_0}\;.
 \end{eqnarray}
where the rotation angle satisfies,
\begin{equation}
     \cot\theta = \tan2\theta_C+\cot\theta_{so}\sec2\theta_C\;.
 \end{equation}
This transformation effectively rotates both the SOC field and the magnetic field such that, at $k_y=0$, the original BdG Hamiltonian (characterized by SOC strengths $\alpha$ and $\beta$, and magnetic-field orientation $\theta_B$) is mapped onto an equivalent Hamiltonian with purely Rashba-type SOC. The resulting effective parameters are given by 
\begin{equation}\label{tilde-alpha}
    \tilde{\alpha}=\lambda\Omega_+\;
\end{equation}
and $\tilde{\beta}=0$, and magnetic field orientation $\tilde{\theta}_B=\theta_B-\theta$. Moreover, in the presence of both Rashba and Dresselhaus SOCs, the Fermi contours (and consequently their intersections with the $k_y$-axis) become anisotropic. As a result, the quantity $k_{so}$ in Eq.~(\ref{kf-pm}) is replaced by
\begin{equation}\label{tilde-kso}
    \tilde{k}_{so}\approx k_\lambda \Omega_-\;,
\end{equation}
with $k_\lambda=m^\ast\lambda/\hbar^2$
In Eqs.~(\ref{tilde-alpha}) and (\ref{tilde-kso}) we have we have introduced the notation,
\begin{equation}
    \Omega_{\pm}=\sqrt{1\pm\sin2\theta_{so}\sin2\theta_C}\;.
\end{equation}

It follows from the above discussion that the free energy, critical currents, and diode efficiency in the presence of both Rashba and Dresselhaus SOC can be estimated using Eqs.~(\ref{Energy})–(\ref{q-cooper-0}), upon performing the parameter substitutions $(\alpha,\beta,k_{so},\theta_B)\rightarrow (\tilde{\alpha},\tilde{\beta},\tilde{k}_{so},\tilde{\theta}_B)$. In particular, to leading order, the diode efficiency can be expressed as
\begin{equation}\label{eta-approx-q}
    \eta \approx (1-\tau)^{1/4}\frac{m^\ast W_N \Omega_-}{\hbar^2 k_F}\lambda\; \tilde{q}
\end{equation}
where the anisotropic Cooper pair momentum is given by
\begin{equation}\label{q-cooper-1}
    \tilde{q}\approx\kappa_B^2\frac{\sin\theta_B\cos\theta_{so}-\sin\theta_{so}\cos(2\theta_C+\theta_B)}{k_F^2\Omega_+}\;.
\end{equation}
In the limit $\theta_{so}=0$ (i.e., $\alpha \neq 0$, $\beta = 0$), this expression reduces to the Rashba-only result [Eq.~(\ref{q-cooper-0})], and the diode efficiency exhibits, to leading order, a sinusoidal dependence on the magnetic field orientation $\theta_B$. Moreover, $\lambda \tilde{q} \propto \mathbf{w}_x \cdot \mathbf{B}$, showing that Eq.~(\ref{q-cooper-1}) is consistent with the phenomenological prediction [Eq.~(\ref{eta-minimal})].

\begin{figure}[t!]
    \centering
    \includegraphics[width=1\columnwidth]{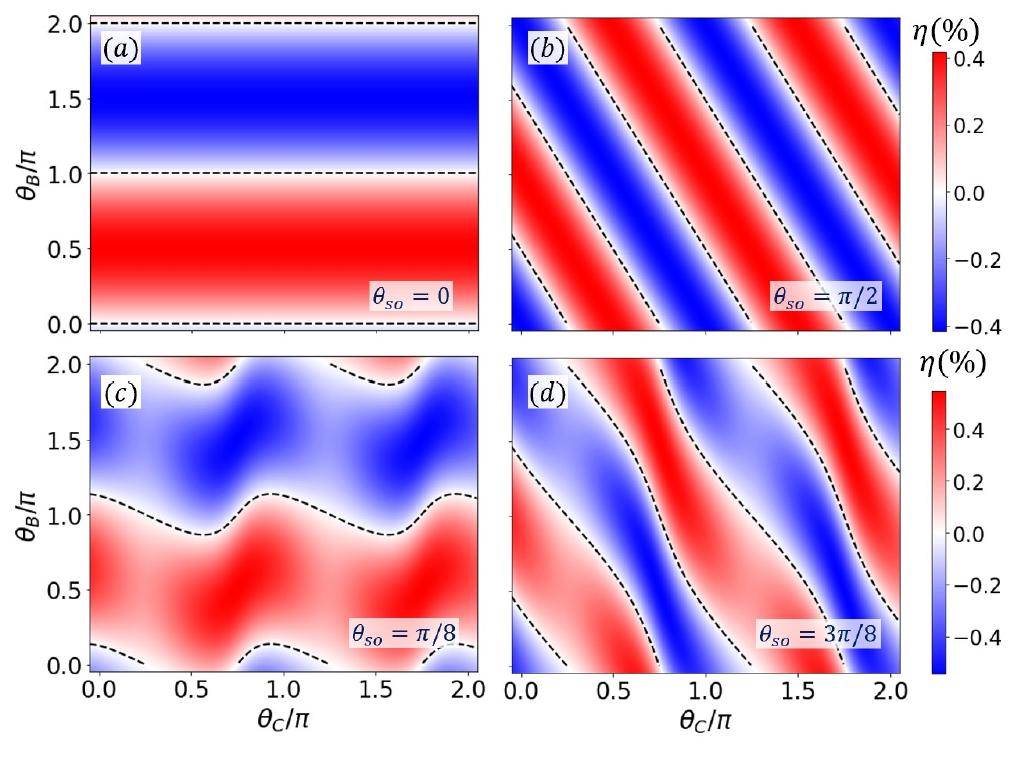}
    \caption{Diode efficiency as a function of the junction orientation $\theta_C$ and the magnetic-field direction $\theta_B$ for SOC angles $\theta_{so}=0$ (a), $\pi/2$ (b), $\pi/8$ (c), and $3\pi/8$ (d). The calculations are performed for an Al/InAs junction using the analytical model defined by Eqs.~(\ref{CPR})–(\ref{qualityfactor}) and (\ref{Energy})–(\ref{Free_energy}). Dashed lines denote the solutions of Eq.~(\ref{thetaB-noSDE}), corresponding to zeros of the Cooper-pair momentum. The magnetic-field strength is set to $B=0.1$~T, and the junction transparency to $\tau=0.99$.}
\label{fig2}
\end{figure}

It follows from Eqs.~(\ref{eta-approx-q}) and (\ref{q-cooper-1}) that, in the presence of both Rashba and Dresselhaus SOCs, deviations from a purely sinusoidal dependence are generally expected. However, for a junction orientation $\theta_C = \pi/4$, the leading-order dependence on $\theta_B$ remains approximately sinusoidal even when both SOC contributions are present. This behavior is in qualitative agreement with the experimental results reported in Ref.~\cite{Schiela2025b:arxiv}.

The effects of the interplay between magnetic and crystalline anisotropies on the diode efficiency, as obtained from the analytical model, are shown in Figs.~\ref{fig2}(a) - (d) for junctions with purely Rashba SOC ($\theta_{so} = 0$), purely Dresselhaus SOC ($\theta_{so} = \pi/2$), Rashba-dominated ($\theta_{so} = \pi/8$, i.e., $\beta \approx 0.4\alpha$) and Dresselhaus-dominated ($\theta_{so} = 3\pi/8$, i.e., $\beta \approx 2.4\alpha$) SOC, respectively, at a magnetic field strength of $B = 0.2$~T. In the purely Rashba case, the diode efficiency is independent of $\theta_C$ [see Fig.~\ref{fig2}(a)]. In contrast, for purely Dresselhaus SOC, it exhibits a pronounced crystalline anisotropy [see Fig.~\ref{fig2}(b)]. Consistent with this behavior, the crystalline anisotropy of the superconducting diode effect (SDE) is weaker in the Rashba-dominated regime than in the Dresselhaus-dominated one, as can be seen by comparing Figs.~\ref{fig2}(c) and (d).

The SDE polarity depends strongly on both $\theta_B$ and $\theta_C$, and for fixed magnetic field strength and junction orientation, it can be reversed by tuning the direction of the in-plane magnetic field. The white traces with superimposed dashed lines in Fig.~\ref{fig2} indicate SDE suppression and correspond to the zeros of the Cooper pair momentum ($\tilde{q} = 0$). According to Eq.~\ref{q-cooper-1}, this condition yields the magnetic field angles
\begin{equation}\label{thetaB-noSDE}
\theta_B = \text{arccot}(\tan 2\theta_C + \cot \theta_{so} \sec 2\theta_C)
\end{equation}
at which the SDE vanishes. The dashed lines in Fig.~\ref{fig2} correspond to the solutions of Eq.~(\ref{thetaB-noSDE}).

Equation~(\ref{thetaB-noSDE}) provides a direct way to extract the SOC ratio $\alpha/\beta$ from SDE measurements. By rotating the in-plane magnetic field, one can identify the orientations $\theta_B^\ast$ at which the SDE is suppressed. Substituting these values into Eq.~(\ref{thetaB-noSDE}) yields an estimate of the SOC ratio:
\begin{equation}
\frac{\alpha}{\beta} = \cot\theta_{so} = \cot\theta_B^\ast \cos 2\theta_C - \sin 2\theta_C.
\end{equation}

\section{Numerical Approach}
\label{sec:numerical}
Complementary to the approximate analytical model, we perform numerical simulations of planar JJs based on Al/InAs and Al/InSb heterostructures, where SOC is typically strong. Unlike the simplified analytical treatment, these simulations capture the SDE in greater detail, including finite-size effects.

We model the system using a tight-binding Bogoliubov–de Gennes (BdG) Hamiltonian defined on a two-dimensional square lattice. The discretization is implemented via finite-difference methods using the Kwant package~\cite{Groth2014:NJP}. The lattice constant is set to $a = 10\,\text{nm}$. The width of the normal region is taken as $W_N = 100\,\text{nm}$, while the superconducting leads are wider, with $W_S = 500\,\text{nm}$. The chemical potential in the superconducting regions is fixed at $\mu_S = 2\,\text{meV}$.

We perform simulations for two representative parameter sets corresponding to Al-proximitized InAs- and InSb-based quantum wells. For the InAs-based junction, which is characterized by predominantly Rashba SOC, we use $m^\ast = 0.033\, m_e$ (with $m_e$ the free electron mass), $g^\ast = -10$, and a proximity-induced pairing potential $\Delta = 0.25\,\text{meV}$. The total SOC strength is set to $\lambda = 16\,\text{meV}\cdot\text{nm}$. In contrast, InSb-based junctions exhibit comparable Rashba and Dresselhaus SOC contributions, making them well suited for exploring crystalline anisotropy effects. In this case, we use $m^\ast = 0.013\, m_e$, $g^\ast = -20$, $\Delta = 0.21\,\text{meV}$, and $\lambda = 16\,\text{meV}\cdot\text{nm}$. For both system types, the spin–orbit angle $\theta_{so}$ is varied to probe different SOC regimes, as discussed in the following section.

In what follows, unless otherwise specified, all calculations are performed using the system parameters listed above with zero junction potential.

\section{Magnetoanisotropic SDE}

In this section, we focus on junctions where only Rashba SOC is present ($\theta_{so}=0$), causing the SDE to become anisotropic with respect to the magnetic-field direction while remaining unaffected by the junction's crystallographic orientation. In this scenario, the SDE vanishes when the magnetic field is perpendicular to the junction ($\theta_B = n\pi$), as shown in Table~\ref{tab-1}, and reaches its maximum when $\mathbf{B}$ is aligned along the junction [$\theta_B = (2n+1)\pi/2$, with $n\in\mathbb{Z}]$.

\begin{figure}[t!]
    \centering
  \includegraphics[width=1\columnwidth]{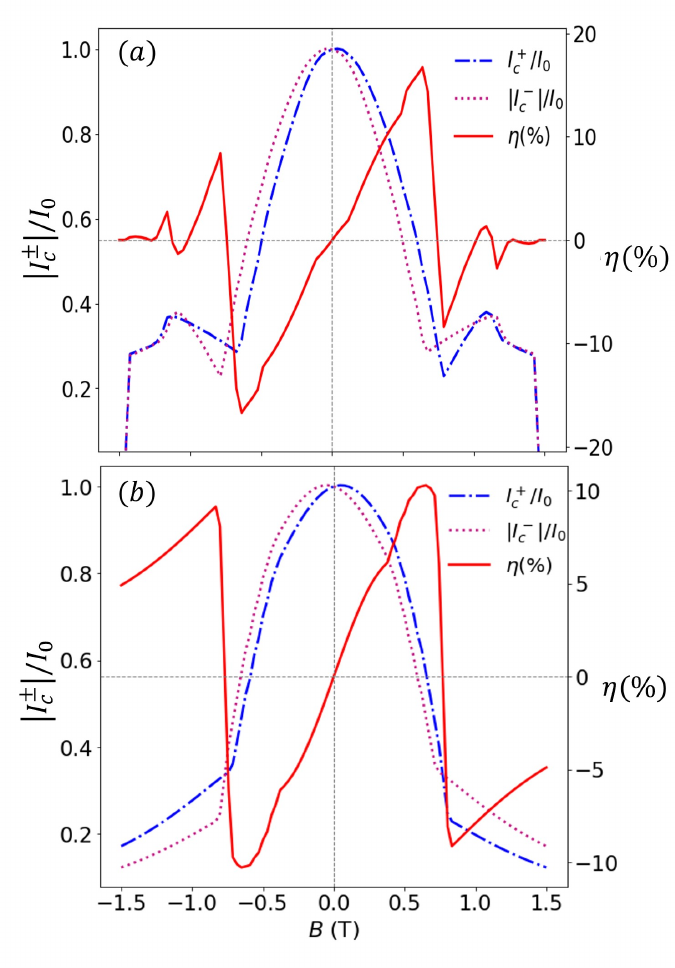}
    \caption{Magnetic-field dependence of the forward ($I_c^+$, dash-dotted line) and reverse ($I_c^-$, dotted line) critical currents, normalized to their zero-field value ($I_0$), for an Al/InAs junction with purely Rashba SOC ($\theta_{so}=0$) and a magnetic field applied along the junction ($\theta_B=\pi/2$). The solid line denotes the corresponding diode efficiency. (a) Numerical simulations and (b) results from the analytical model (with $\tau=0.99$ and $g^\ast=-40$).}\label{fig3}
\end{figure}

Figure~\ref{fig3} shows how the forward and reverse critical currents, along with the diode efficiency, depend on the magnetic field strength for an Al/InAs JJ at $\theta_B = \pi/2$. Panels (a) and (b) correspond to the numerical simulations and the analytical approach, respectively. The behavior of the critical currents and the diode efficiency is qualitatively consistent with recent experimental observations \cite{Lotfizadeh2024:CP}. In particular, beyond the trivial SDE sign reversal near zero field, additional polarity changes appear at higher fields (around $\pm 0.75$~T), close to the fields at which the junction in its equilibrium ground state would undergo $0$–$\pi$ transitions \cite{Costa2023:NN,Lotfizadeh2024:CP}. We note that, while the numerical simulations include a magnetic field applied throughout the entire system, the analytical model assumes the field to be finite only in the N region for simplicity. As a consequence, a larger effective $g$-factor ($g^\ast = -40$) is required in the analytical model to reproduce the field values ($B \approx \pm 0.75$~T) at which the polarity changes associated with the $0$–$\pi$ transitions occur, as obtained in the numerical simulations with $g^\ast = -10$.

\begin{figure}[t!]
    \centering
  \includegraphics[width=1\columnwidth,height=0.5\textwidth]{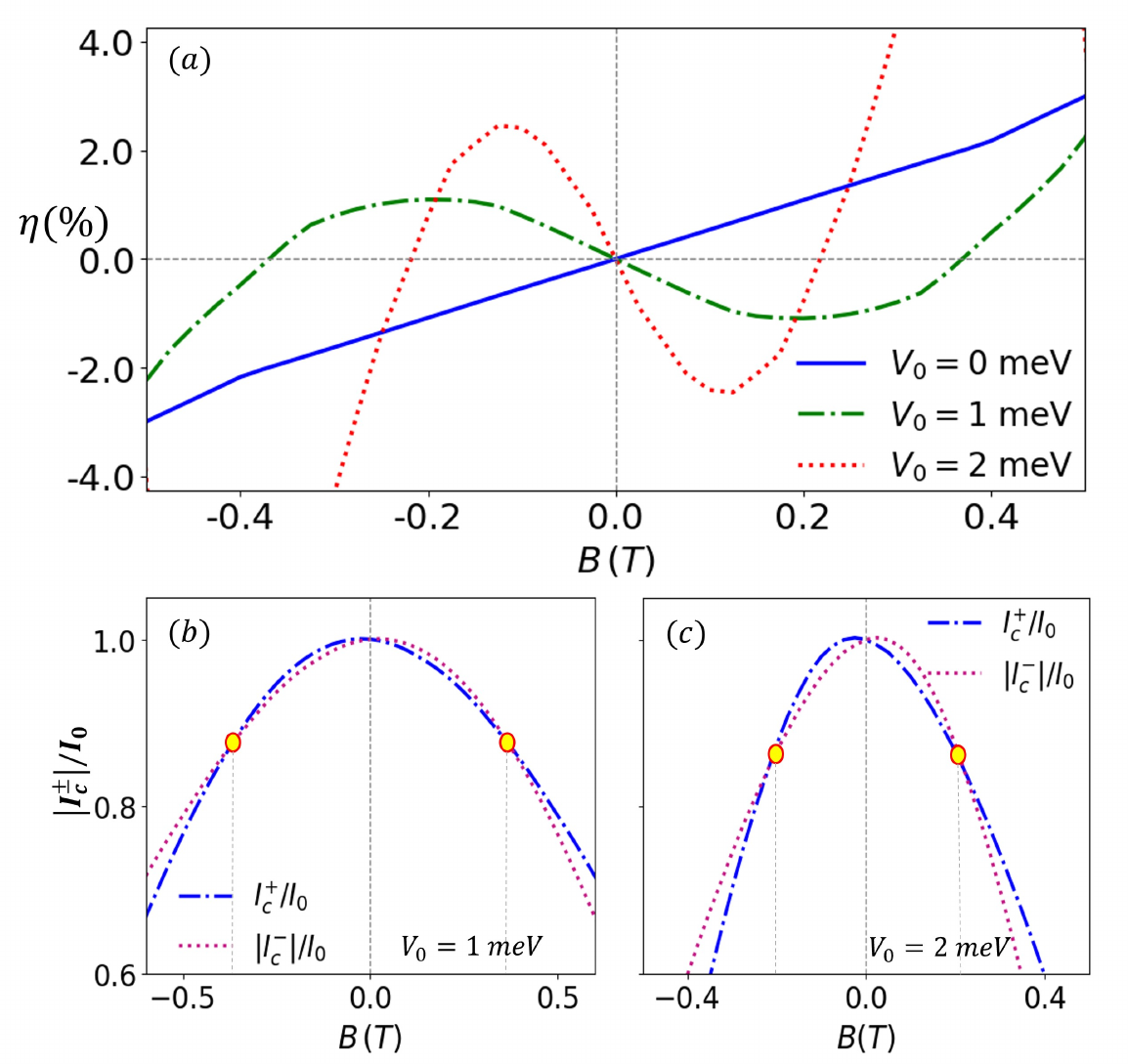}
    \caption{(a) Magnetic-field dependence of the diode efficiency in an Al/InAs junction with purely Rashba SOC ($\theta_{so}=0$), shown for different values of the junction potential $V_0$. Panels (b) and (c) display the corresponding magnetic-field dependence of the forward (dash-dotted line) and reverse (dotted line) critical currents. Yellow markers indicate the crossing points of the critical currents at $B \approx \pm 0.37$~T and $B \approx \pm 0.22$~T, which coincide with the SDE polarity reversals shown in panel (a) for $V_0 = 1$~meV and $V_0 = 2$~meV, respectively. 
    The numerical simulations correspond to a junction with $W_N = 100$~nm and $W_S = 40$~nm, with the magnetic field oriented perpendicular to the current flow (i.e., $\theta_B = \pi/2$). 
    }\label{fig4}
\end{figure}

Interestingly, SDE sign reversals can also occur at lower fields when the potential $V_0$ is properly tuned, which can be achieved using a top gate on the N region of the junction. This behavior has recently been observed experimentally \cite{Schiela2025b:arxiv} and attributed to the interplay between linear and cubic Dresselhaus SOC and gate-induced modulations of the Rashba SOC strength. However, our numerical calculations indicate that SDE polarity switches at low fields can also arise in junctions with narrow superconducting contacts solely from gate-induced changes in the junction potential, even in the absence of Dresselhaus SOC. This is illustrated in Fig.~\ref{fig4}(a), which shows the diode efficiency as a function of magnetic field strength for an Al/InAs junction with only Rashba SOC and $\theta_B=\pi/2$, at three different values of the junction potential. The figure illustrates how tuning $V_0$ from 0 to $1$~meV and $2$~meV modifies the magnetic-field dependence of the diode efficiency, leading to the emergence of SDE polarity reversals at lower fields. As shown in Figs.~\ref{fig4}(b) and (c), which display the magnetic-field dependence of the critical currents, these low-field polarity reversals occur when the forward and reverse critical currents intersect due to differences in the curvature of their $B$ dependence. Importantly, this mechanism is not related to $0$–$\pi$ transitions, which typically arise at higher magnetic fields (see Fig.~\ref{fig3}).

\begin{figure}[t!]
    \centering
  \includegraphics[width=1\columnwidth,height=0.435\textwidth]{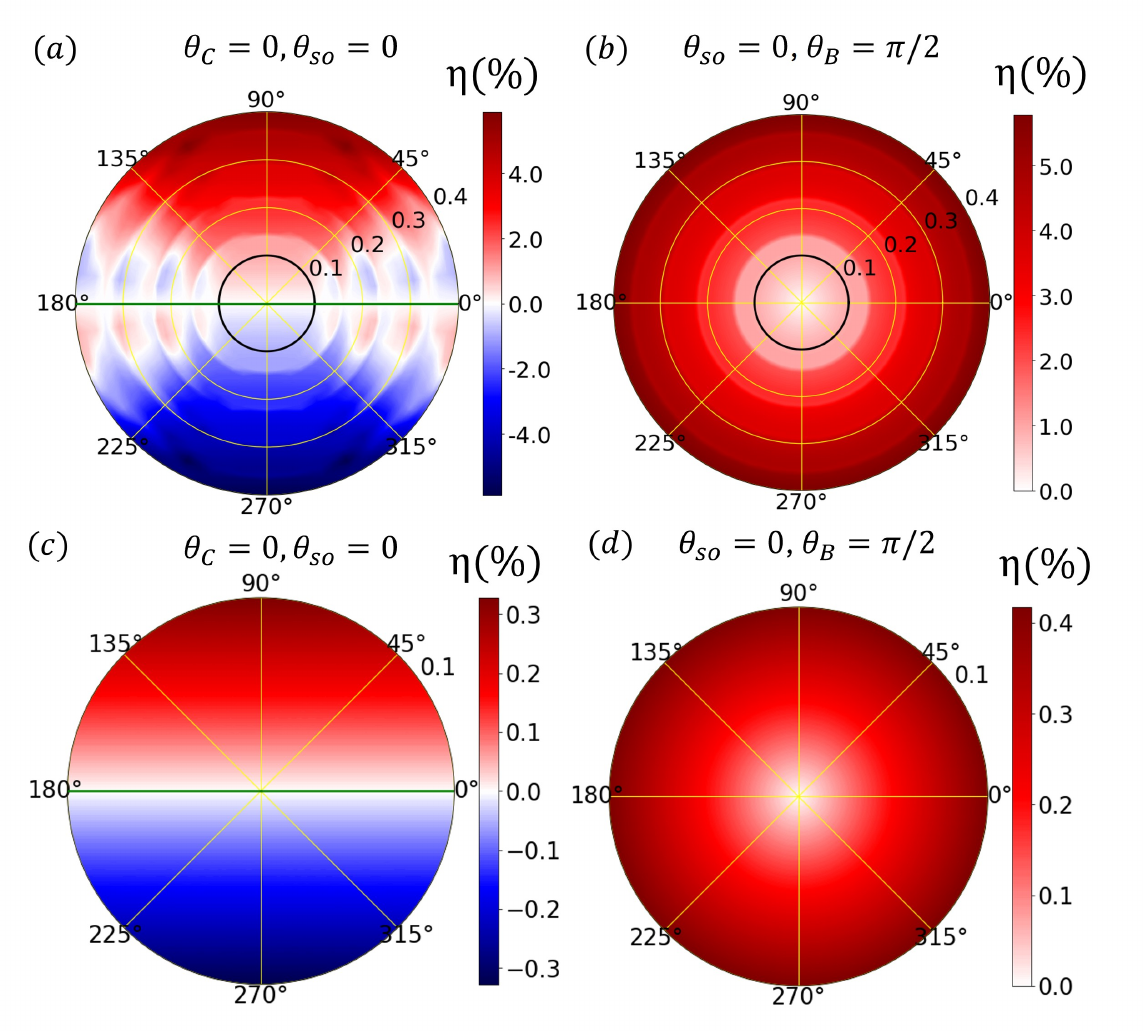}
    \caption{Magnetoanisotropy of the SDE in an Al/InAs junction with purely Rashba SOC ($\theta_{so}=0$). (a) Diode efficiency from numerical simulations as a function of the magnetic-field angle $\theta_B$ (polar angle) and the field strength $B$ (radial coordinate) for $\theta_C=0$. (b) Diode efficiency from numerical simulations as a function of the junction orientation $\theta_C$ (polar angle) and the field strength $B$ (radial coordinate) for $\theta_B=\pi/2$. Panels (c) and (d) show results from the analytical model (with $\tau=0.99$) corresponding to (a) and (b), respectively, within the low-field region indicated by the black circles in (a) and (b). Green lines denote field directions for which symmetry-imposed suppression of the SDE occurs independently of the field strength.}\label{fig5}
\end{figure}

To illustrate the magnetoanisotropy of the SDE, we computed the diode efficiency as a function of the magnetic field angle $\theta_B$ (polar angle) and the field strength $B$ (radial coordinate) for $\theta_{so}=0$. The results, corresponding to the numerical simulations and the analytical approach, are shown as polar plots in Figs.~\ref{fig5}(a) and (c), respectively. The analytical results agree well with the numerical simulations at weak magnetic fields. The green lines indicate field directions ($\theta_B=0,\pi$) for which the SDE is suppressed regardless of field strength, confirming the predictions in Table~\ref{tab-1}. Additional zeros of the diode efficiency, which depend on $B$ and are therefore not symmetry-related, appear in the numerical results shown in panel (a) and are attributed to finite-size effects and higher-order contributions. For completeness, we show in Figs.~\ref{fig5}(a) and (b) the dependence of $\eta$ on the junction crystallographic orientation $\theta_C$ and magnetic field strength for $\theta_B = \pi/2$. As expected, both the numerical [panel (b)] and analytical [panel (d)] results exhibit rotational symmetry, confirming the lack of crystalline anisotropy when Dresselhaus SOC is not present.

\begin{figure}[t!]
    \centering
  \includegraphics[width=1\columnwidth,height=0.42\textwidth]{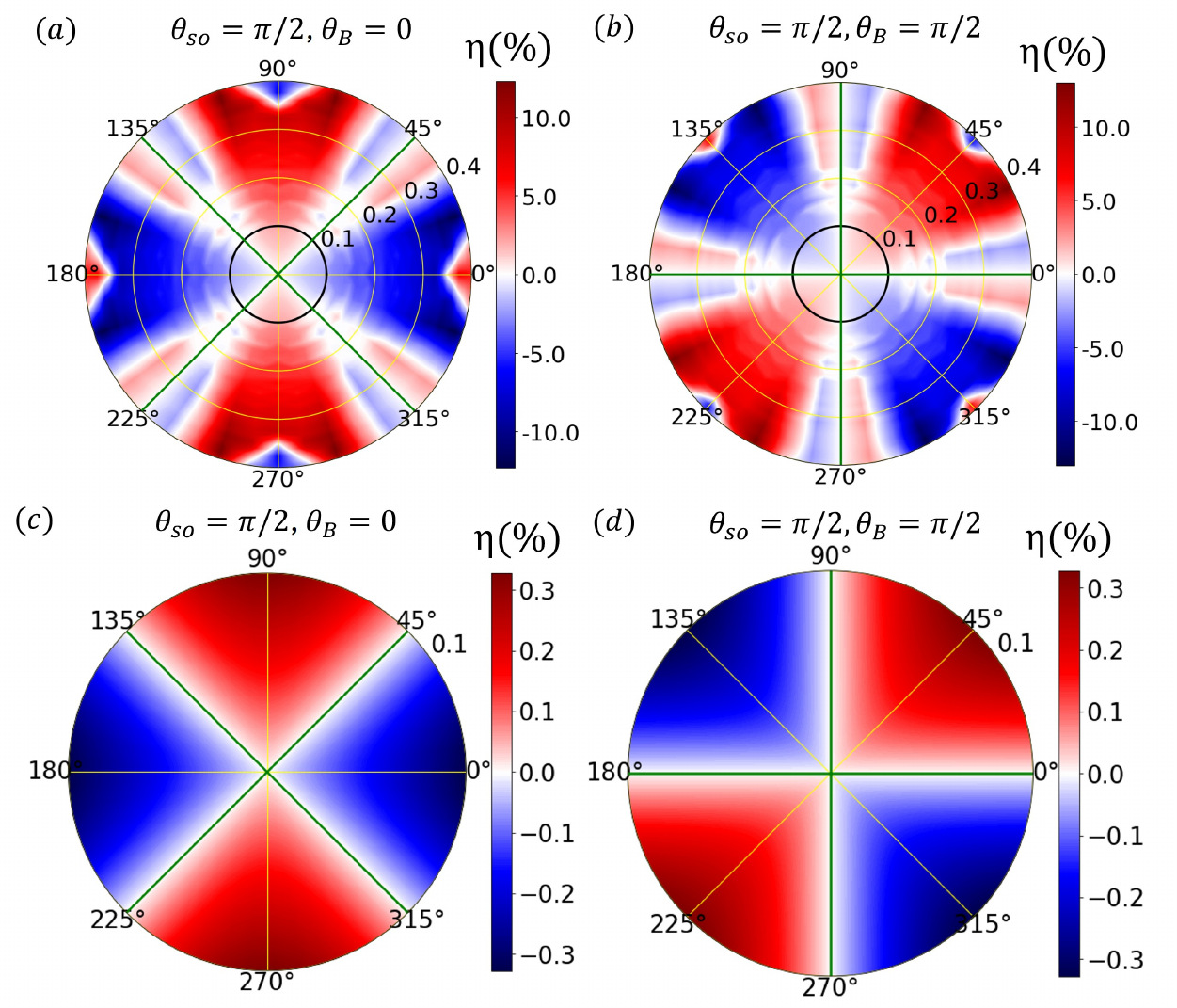}
    \caption{Crystalline anisotropy of the SDE in an Al/InSb junction with purely Dresselhaus SOC ($\theta_{so}=\pi/2$). (a) Diode efficiency from numerical simulations as a function of the junction orientation $\theta_C$ (polar angle) and the field strength $B$ (radial coordinate) for a magnetic field applied along the current direction ($\theta_B=0$). (b) Same as in (a), but for a magnetic field perpendicular to the current direction ($\theta_B=\pi/2$). Panels (c) and (d) show results from the analytical model (with $\tau=0.99$) corresponding to (a) and (b), respectively, within the low-field region indicated by the black circles in (a) and (b). Green lines denote junction orientations for which symmetry-imposed suppression of the SDE occurs independently of the field strength.}
    \label{fig6}
\end{figure}

\section{ Magneto- and Crystalline-Anisotropic SDE} 

Unlike Rashba SOC, Dresselhaus SOC is not invariant under in-plane rotations and therefore introduces, in addition to magnetoanisotropy, a crystalline anisotropy whereby the SDE depends on the current direction, which is set by the junction crystallographic orientation.

We first consider a junction with only Dresselhaus SOC ($\theta_{so} = \pi/2$). The diode efficiency, computed numerically as a function of the junction crystallographic angle $\theta_C$ (polar angle) and the magnetic field strength (radial coordinate), is shown in Figs.~\ref{fig6}(a) and (b) for magnetic field orientations $\theta_B = 0$ and $\theta_B = \pi/2$, respectively. In the low-field regime, the analytical results, presented in panels (c) and (d), are in good agreement with the numerical simulations.  A clear twofold crystalline anisotropy is observed for both selected magnetic field orientations. As the magnetic field increases, higher-order contributions become significant, leading to more complex and richer anisotropic patterns, as shown in panels (a) and (b). For $\theta_B = 0$ ($\theta_B = \pi/2$), the SDE is suppressed at junction orientations $\theta_C = (2n+1)\pi/4$ ($\theta_C = (2n+1)\pi/2$), indicated by green lines, and in agreement with the predictions of Table~\ref{tab-1}. Figure~\ref{fig6} also shows that, in the presence of only Dresselhaus SOC, the SDE is maximized when the magnetic field is aligned with the current direction. This contrasts with the Rashba-only case, where the optimal SDE occurs when the magnetic field is oriented along the junction, i.e., perpendicular to the current flow.

\begin{figure}[t!]
    \centering
  \includegraphics[width=1\columnwidth,height=0.46\textwidth]{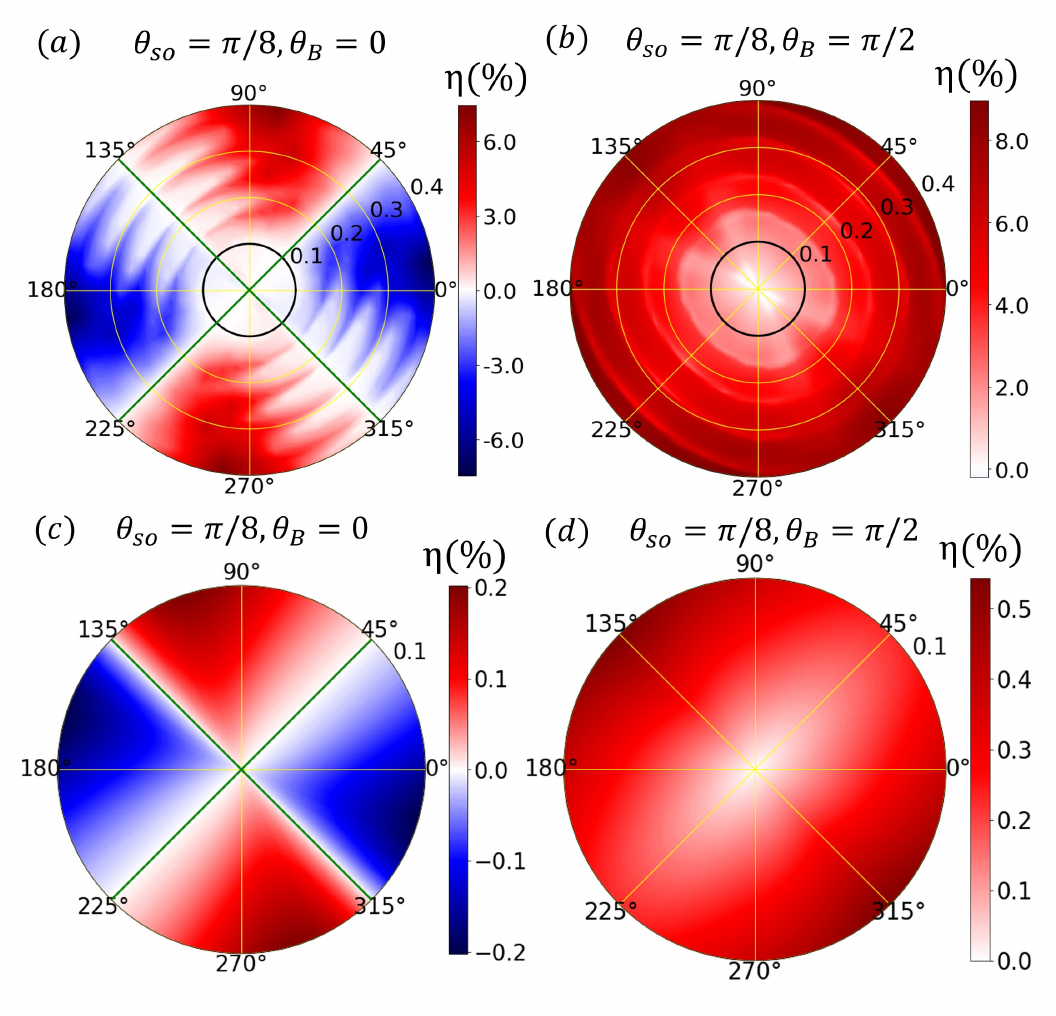}
    \caption{Crystalline anisotropy of the SDE in a Rashba-dominated Al/InAs junction with coexisting Rashba and Dresselhaus SOCs. (a)–(d) Same as in Figs.~\ref{fig6}(a)–(d), but for $\theta_{so}=\pi/8$ ($\beta \approx 0.4\alpha$).}
    \label{fig7}
\end{figure}

The behavior of the diode efficiency for a junction with both Rashba and Dresselhaus SOC, with spin-orbit angle $\theta_{so} = \pi/8$ ($\beta \approx 0.4\alpha$), is shown in Fig.~\ref{fig7}. The top and bottom rows correspond to the numerical simulations and the low-field analytical approach, respectively. The diode efficiency exhibits Crystalline anisotropy for both $\theta_B = 0$ and $\theta_B = \pi/2$.

Despite the Rashba SOC strength being nearly twice that of the Dresselhaus SOC, the crystalline anisotropy at $\theta_B = 0$ closely resembles the Dresselhaus-only case (compare the left columns of Figs.~\ref{fig6} and \ref{fig7}). This occurs because, for $\theta_B = 0$, the magnetic field is aligned with the current direction, thereby reducing the influence of Rashba SOC. Consistent with the symmetry conditions in Table~\ref{tab-1}, for $\alpha \neq 0$, $\beta \neq \alpha$, and $\theta_B = 0$, the SDE is suppressed at junction orientations $\theta_C = (2n+1)\pi/4$, highlighted by green lines in Figs.~\ref{fig7}(a) and (c).

In contrast, for $\theta_B = \pi/2$, the magnetic field is oriented along the junction (i.e., perpendicular to the current flow), enhancing the role of Rashba SOC. This results in a crystalline anisotropic SDE pattern that more closely resembles the Rashba-only case, while still retaining anisotropy. In this configuration, the symmetry conditions in Table~\ref{tab-1} cannot be satisfied for any $\theta_C$, and therefore no junction orientation leads to symmetry-imposed suppression of the SDE.

\begin{figure}[t!]
    \centering
  \includegraphics[width=1\columnwidth,height=0.44\textwidth]{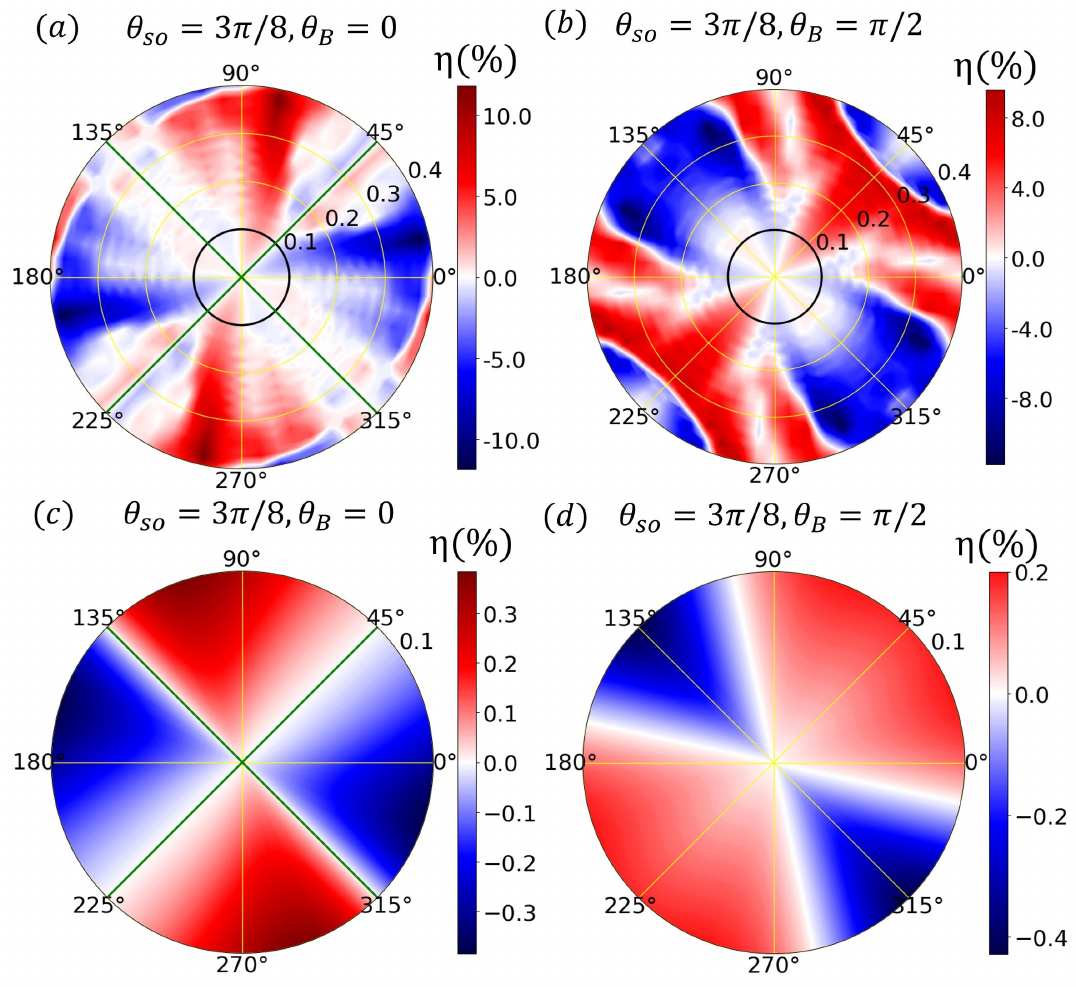}
    \caption{Crystalline anisotropy of the SDE in a Dresselhaus-dominated Al/InSb junction with coexisting Rashba and Dresselhaus SOCs. (a)–(d) Same as in Figs.~\ref{fig6}(a)–(d), but for $\theta_{so}=3\pi/8$ ($\beta \approx 2.4\alpha$).}
    \label{fig8}
\end{figure}

Complementary to Fig.~\ref{fig7}, Fig.~\ref{fig8} illustrates the crystalline anisotropy of the SDE in a Dresselhaus-dominated Al/InSb junction with $\theta_{so}=3\pi/8$ ($\beta \approx 2.4~\alpha$). The top and bottom rows present the numerical and low-field analytical results, respectively. The symmetry-imposed zeros of the SDE—indicated by the green lines in Figs.~\ref{fig8}(a) and (c)—are consistent with condition (3) in Table~\ref{tab-1}. Note, however, that the SDE zeros in Figs.~\ref{fig8}(b) and (d) depend explicitly on the magnetic field strength and are therefore not symmetry-protected.

Notably, in the Dresselhaus-dominated regime, polarity reversals of the SDE can occur at low magnetic fields for specific crystallographic orientations, even in the absence of a junction potential. This effect is particularly pronounced when the magnetic field is aligned with the junction ($\theta_B=\pi/2$), as clearly illustrated in Fig.~\ref{fig8}(b), where SDE polarity reversals appear at $B \approx 0.13$~T and $B \approx 0.31$~T for junction orientations $\theta_C = (4n+3)\pi/4$ and $\theta_C = (4n+1)\pi/4$, respectively. These results demonstrate that SDE sign reversals at low magnetic fields can arise not only from electrostatic gating but also from the geometric configuration of the magnetic field and junction orientation.

\begin{figure}[t!]
    \centering
  \includegraphics[width=1\columnwidth]{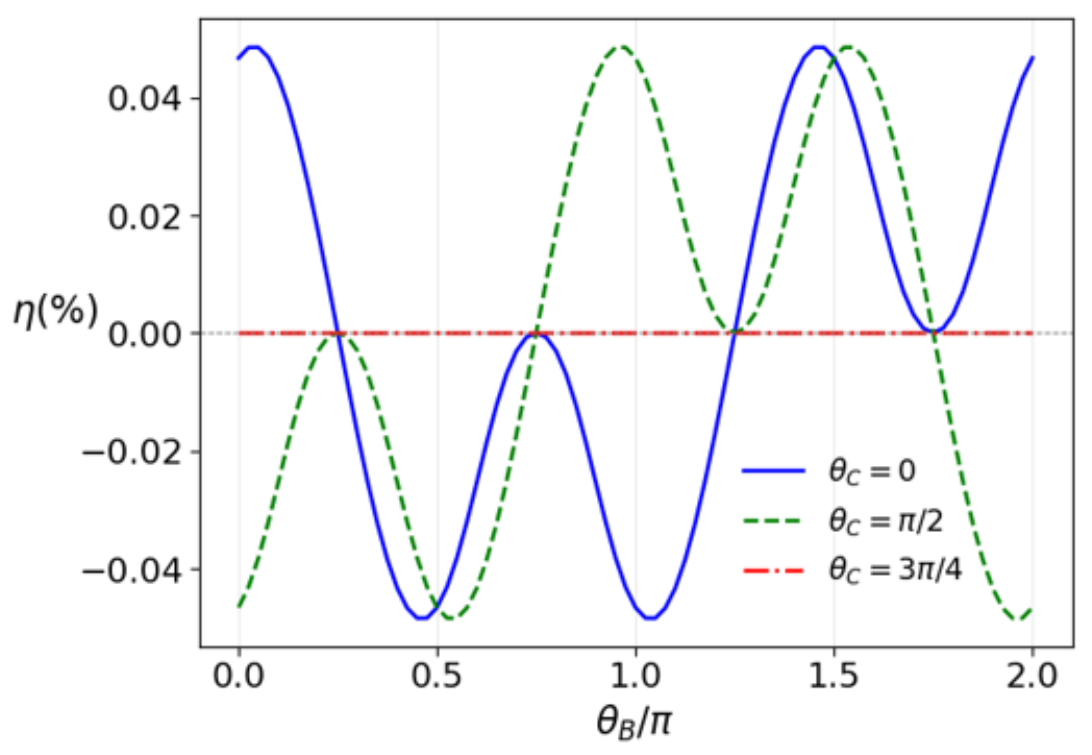}
    \caption{Diode efficiency from numerical simulations as a function of the magnetic-field angle $\theta_B$ for Al/InSb junctions at $B=0.1$~T, with $\alpha=\beta$ ($\theta_{so}=\pi/4$), and junction orientations $\theta_C=0$ (blue solid line), $\theta_C=\pi/2$ (green dashed line), and $\theta_C=3\pi/4$ (red dash-dotted line). Zeros of $\eta$ accompanied by sign reversals satisfy condition (4) in Table~\ref{tab-1}, while those occurring without sign changes satisfy condition (5).}
\label{fig10}
\end{figure}

As an illustration, Fig.~\ref{fig10} shows the dependence of $\eta$ on the magnetic field orientation for a junction with $\alpha = \beta$ ($\theta_{so} = \pi/4$), for different junction orientations and $B = 0.1$~T. The figure reveals a distinctive signature: zeros associated with the fulfillment of Eq.~(\ref{u-sym}) [condition (4) in Table~\ref{tab-1}] are accompanied by polarity reversals of the SDE, whereas those related to gauging out the SOC [condition (5) in Table~\ref{tab-1}] do not involve a change of sign. A particularly notable case occurs for $\alpha = \beta$ and $\theta_C = 3\pi/4$, where the SDE vanishes for any in-plane orientation of the magnetic field.

\section{Summary}

We theoretically investigate the magnetic and crystalline anisotropies of the SDE in proximitized planar JJs with Rashba and Dresselhaus SOC under an in-plane magnetic field. A symmetry analysis reveals geometric constraints on field and crystallographic orientations for which the SDE is suppressed, independent of field strength. These predicted conditions provide experimentally testable signatures to assess whether the interplay between SOC and Zeeman interaction underlies the SDE in semiconductor-based planar JJs.

We develop a phenomenological model showing that the diode efficiency depends on the relative alignment between the spin–orbit and magnetic fields. This behavior is confirmed in the narrow-junction, low-field regime by an analytical model, which links the anisotropy of the diode efficiency to the SOC-induced deformation of the Fermi contours and the anisotropy of the Cooper pair momentum. These results are supported by tight-binding simulations of the Bogoliubov–de Gennes equation, and are consistent with recent experimental trends. Notably, the simulations show that electrostatic gating can induce SDE polarity reversals in the low-field regime even with only Rashba SOC, in agreement with recent experiments \cite{Schiela2025b:arxiv}. Additional polarity reversals are predicted for specific magnetic field orientations, junction geometries, and SOC ratios $\alpha/\beta$.

Overall, our results clarify how SOC and Zeeman coupling generate anisotropic nonreciprocal transport and provide guidance for experimentally probing the mechanisms behind the SDE through its predicted anisotropic diode response.

\begin{acknowledgments}
We thank William F. Schiela and Javad Shabani for helpful discussions.
\end{acknowledgments}

\bibliographystyle{apsrev4-2}
\bibliography{Reference}

\end{document}